\documentclass[useAMS,usenatbib,usegraphicx]{mn2e}

\title[The Galactic IMF: origin in the combined mass distribution functions of dust grains 
and gas clouds]{The Galactic IMF: origin in the combined mass distribution functions of dust grains 
and gas clouds}

\author[E. Casuso and J. E. Beckman]{E. Casuso$^{(1,3)}$\thanks{E-mail:
eca@iac.es} and J. E. Beckman$^{(1,2,3)}$\footnotemark[1]\\\
$^{1}$Instituto de Astrofisica de Canarias, 38205, La Laguna, Tenerife, Spain\\
$^{2}$C.S.I.C., Madrid, Spain\\
$^{3}$Department of Astrophysics, University of La Laguna,
Tenerife, Spain }
\begin{document}

\date{Accepted   . Received  ; in original form } 

\pagerange{\pageref{firstpage}--\pageref{lastpage}} \pubyear{2011} 

\maketitle

\label{firstpage}

\begin{abstract}
We present here a theoretical model to account for the stellar IMF as a result of the composite behaviour of the gas and
dust distribution functions. Each of these has previously been modelled and the models tested against observations. The
model presented here implies a relation between the characteristic size of the dust grains and the characteristic final
mass of the stars formed within the clouds containing the grains, folded with the relation between the mass of a gas cloud
and the characteristic mass of the stars formed within it. The physical effects of dust grain size are due to equilibrium 
relations between the efficiency of grains in cooling the clouds, which is a falling function of grain size, and the efficiency
of grains in catalyzing the production of molecular hydrogen, which is a rising function of grain size. We show that folding in 
the effects of grain distribution can yield a reasonable quantitative account of the IMF, while gas cloud mass function alone
cannot do so.
\end{abstract}


\section{Introduction}

When we observe the mass distribution function of stars in a galaxy we obtain the present day mass function (PDMF)
which is not the mass function of stars at their birth (the initial mass function, IMF) but the
distribution resulting from the accumulation of stars since the first star formation occurred in the Galaxy.
In this process, whether the star formation is continuous or sporadic, the most massive stars with their
short lifetimes are more quickly eliminated, and the masses of all the stars are reduced secularly by mass loss,
while the stellar population may be augmented by the arrival of stars from outside the Galaxy in accretion events
with dwarf galaxies. The local IMF is a basic function in the study of the chemical evolution of our Galaxy, as it
is the measure of the number of stars formed in a given mass interval in the solar neighborhood \citep{b25,b26,b31}.
The origin of the IMF is a fundamental problem in the whole of astrophysics because it determines 
the photometric properties of galaxies, and the dynamical and chemical evolution of their interstellar media. \citet{b24} was the
first to note that there appears to be a power-law relationship between the number of stars observed in the Galactic field and their masses,
and it was his work which gave rise to our concept of the IMF.

In order to try to account for the form of the IMF one needs to know the mechanism of the process of star
formation. We understand that this process begins with the formation of giant molecular clouds within which, given
the right conditions of temperature and density (the Jeans conditions \citet{b13}) local gravitational collapse gives
rise to stars. It is also well understood that the presence of dust grains as catalysts is an essential condition
for the conversion of atomic to molecular hydrogen, so that they are necessary for the presence of the giant
molecular clouds. It thus appears reasonable that in order to investigate the factors which explain the stellar
IMF one needs to take into account not only the gas cloud mass distribution function, but also the distribution
function of the dust grain sizes. The relationship between the gas and star formation is obvious since stars form
from the gas clouds, but the connection between the dust and star formation is less direct. As we have just noted,
the presence of dust, even in small quantities, makes a major difference to the conversion rate of atomic to
molecular hydrogen \citet{b28}, but dust also affects the heating cycle in the ISM, which affects the tendency
to form the cold dense clouds needed for star formation.

Arguments based on gravitational instability and on observations of molecular gas reveal that the low-temperature
(T $\sim$ 10 K) high-density ($n$ $\geq$ 40 cm$^{-3}$) cores in giant molecular clouds (GMC's) are the natural
sites for stars to form. Although individual GMC's are well resolved in the Milky Way and in our local group (\citet{b4},
and references therein), it is from CO observations of nearby spirals which show most clearly that star formation
 occurs in regions dominated by molecular gas \citep{b33,b14,b1}.

The present article, whose aim is to bring out the connection between cloud mass distributions, dust grain size
distributions, and the stellar mass distribution, is organized as follows: in Section 2 we present the version we
will be using of the stellar IMF derived from published observations by a number of authors,
 in Section 3 we develop three models in which the observed mass functions for the Giant Molecular Clouds in the Galaxy
 will be used, in conjunction with the interstellar dust grain size distribution, to predict the IMF: the zero model (which we will combine
 with the three models of gas clouds to obtain the IMF) is our numerically
 derived dust grain size distribution function taken from \citet{b7} (CB10) which fits well the observed carbonate and silicate distributions, and takes into account both
 the production and the destruction of grains, the latter by grain-grain collisions in the ISM. The first model uses the numerically derived molecular cloud mass
 distribution function, taken from \citet{b6} (CB07) in which clouds may coagulate to form bigger clouds or may be disrupted in collisions, depending on their masses,
 temperatures, densities and relative velocities, while a fraction condense to form stars. The second model uses an analytically derived molecular cloud
  mass distribution function taken from \citet{b5} (CB02)
 where we solve the stochastic differential Langevin equation in a context of existing barriers due to box-effect. The third model uses a fully
 analytic approximation based on the assumption of balance between the thermal emission by dust and dust heating due to collisions with gas molecules, and is based on
 the CB02 approximation, 
 in Section 4 we compare the predictions of the models with the observationally derived IMF taken form several authors, and in
 Section 5 we present our conclusions.
 
\section{The observationally derived IMF's}

As the observational references required to test our models we have used three observational studies of the local stellar mass distribution at birth (i.e.
the IMF). The first reference is \citet{b18} who derive two possible IMF's, one of which assumes a constant star formation rate (SFR) and
the other assumes a variable SFR derived from observations of late-type stellar populations on the main sequence. Both of these versions of the IMF used
the present day stellar mass function (PDMF) by \citet{b16}. We have plotted these in Fig. 1, where we can see the difference in the slopes of the two curves due to
the well understood trend of stars to disappear from the PDMF at a greater rate the higher the initial mass, and the different effect this produces using
different histories for the SFR. The change in slope required to produce the IMF from the PDMF is greater with the variable SFR, because this includes a
relatively recent observed peak in the local Galactic SFR \citep{b18,b23}.

We have used two other references for the Galactic IMF, one obtained by \citet{b17} which takes into account constraints from local star count data, and
 \citet{b27} compilation of MF power law indices for young clusters and OB associations, and which shows different slopes for different mass ranges, and another,
  the semi-empirical Galactic IMF obtained recently by \citet{b21}, both of them also shown in Fig. 1.

\begin{figure}
\includegraphics[width=90mm]{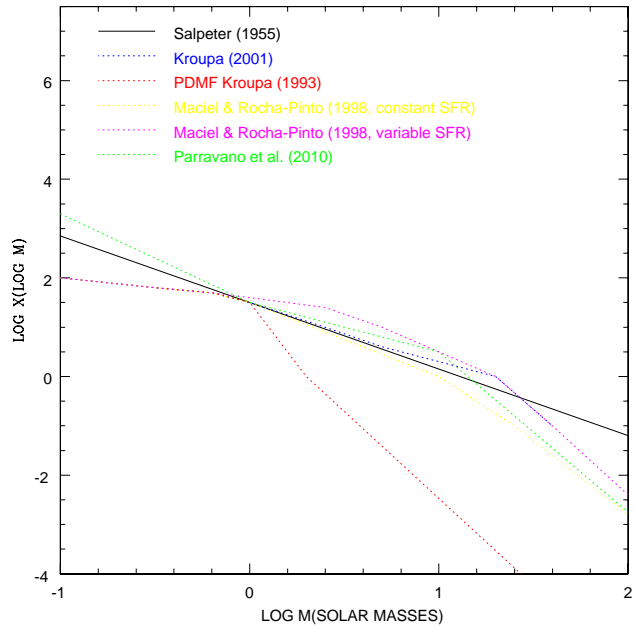}
\caption{Observationally based stellar mass functions for the solar neighbourhood, compared with the classical
Salpeter IMF. {\bf X(log M) is the number of stars per logarithmic mass interval.}}
\end{figure}
 
\section{The theoretical models, folding the gas and dust distribution functions}

\subsection{Semi-analytical treatments}

 \subsubsection{The numerically derived dust grain size distribution function (GSDF)}

Knowing that ISM dust grains are generally formed in the outer atmospheres of stars in the late stages of their evolution (mainly red giants, asymptotic giant
branch stars, and SNe) and that grain lifetimes can be estimated as less than 100 Myr, we have taken an observational version of the local stellar birthrate to
calculate the radius distribution of grains surviving today. To do so, we computed the mass distributions of stars present over this period assuming a standard
IMF, the observed stellar birthrate function, and stellar lifetimes as a function of their masses, and folded in the initial distribution of dust grain sizes at
each epoch, according to a simple prescription of grains produced in different stellar mass ranges. The differences between the size distribution of grains
produced in the older, low mass stellar population and those produced in younger, high mass stars include the tendency of the smaller grains to be
preferentially destroyed by shocks in the post-SN environment. Under these assumptions, the major peak in the grain size distribution comes from dust produced
in the younger more massive stars, and the three peaks in the interstellar dust grain size distribution function (GSDF) of carbonaceous grains correspond to
the three maxima in the local SFR, as detailed in \citet{b23}. This correlation between the three main star formation events in the Galaxy and
the three peaks observed for the sizes of grains appears clear. For silicates, the observed GSDF shows only one peak, close to the biggest peak in the
distribution observed for carbonaceous grains. This increases the plausibility of the scenario in which the peak for carbonaceous grains at the smallest radii
is associated with those grains produced during an interval of some 10 Myr near the late-stages stars of lower masses (near 1 M$_\odot$ arising from the earliest major
peak in the SFR) because in that case the velocities of grains in the expanding shells are lower than in stars with higher masses so that the time interval for
collisions among grains is greater so the probability of shattering is higher and the mean sizes of the grains are lower. For the second peak, we have grains
coming from stars of intermediate age (intermediate mass coming from the second peak in the SFR) where the velocities of grains are intermediate and so some of
the larger grains can escape from the high density shells where the shattering can break them into smaller grains. Finally the third peak, with the largest
grains, can come from SNe where the expansion velocities are so high that the largest grains can escape from the shells into the ISM where the densities are so
low that the collisions occur relatively infrequently. The observed difference in the slope of the distribution function for carbonaceous (close to 0) and
silicate grains (close to 1) may be due to the mechanism proposed by \citet{b32} for suprathermal grains, whereby the mislaignment with respect to the
interstellar magnetic field, may be stronger for carbonaceous dust than for silicate dust. This would lead to greater collisional effects for the carbonaceous
grains, changing the initial slope to that observed. The global features of the GSDF are rather well reproduced by the model in which we use only the dust
sizes yielded by the stellar production process and the accompanying modifications due to shattering in the local environment of the late stages of the stars,
together with a stellar birthrate function from observations. However, a closer fit to the carbonaceous grain distribution at the low radius end is obtained if
we also include the modeling of longer term grain-grain collisions processes in the ISM (CB10). In Fig. 2 we can see the complete model (full line)
and the comparison with data.

\begin{figure}
\includegraphics[width=90mm]{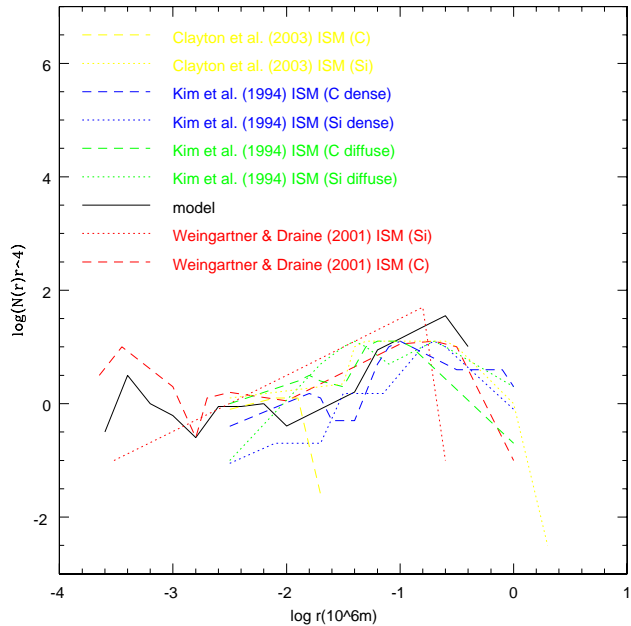}
\caption{Model of the GSDF produced by combining the stellar grain production curve with the effects of size modification
due to collisions in the circumstellar environment. Observational data sets are shown for comparison.}
\end{figure}

 \subsubsection{The numerically derived molecular cloud gas mass distribution function (MCGMF).}
As one can see in CB07 we parameterize the distribution mass function of gas clouds at a given epoch as a function of their
initial distributions of density, temperature, velocity, and mass. The position and velocity distributions are three-dimensional.
We assume different sets of initial distribution functions, (Gaussian, flat, power-law) for temperature, densities, masses, and
velocities of an initial sample of $\sim$1000 gas clouds, either pressure bound or gravitationally bound, evolving within a
confinement volume specified as a cubic box, and arranged initially as knots in a cubic grid. Because our best fit to data is
found for Gaussian distributions, and for spheroidal gas clouds we have taken that case for the present work. Each
cloud has an effective volume that depends on its mass and density via $V{\simeq}{\frac{M}{\rho}}$. After a given evolution time,
with a given initial set of velocities, we consider that two of the clouds have collided if the distance between their centers is
less than or equal to the sum of their radii $r$, which are calculated using $r = \left(\frac{3M}{4{\pi}{\rho}}\right)^{1/3}$. To
take particular values for the main physical parameters (density, temperature, velocity, and mass) we take the expression:
\begin{equation}
f(x) = \frac{1}{\sigma(2\pi)^{1/2}}e^{-\left(\frac{X-X_{0}}{2^{1/2}\sigma}\right)^{2}}
\end{equation}
where $f(x)$ is the probability distribution function of a variable $X$ centred on a value $X_{0}$  and with a dispersion $\sigma$.
We have inverted these functions using the formula:
\begin{equation}
X = X_{0} + {\Delta}X(-2\sigma^{2}log[f_{1}(x)])cos[2{\pi}f_{2}(x)]
\end{equation}
where $f_{1}$ and $f_{2}$ are two random values of the distribution, and ${\Delta}X$ is the range of values permitted for the variable $X$.
To take into account the phenomenon of gas cloud collapse to form stars, we assume that those gas clouds that have attained the Jeans mass
disappear and we count them as new stars (however, if these stars have masses big enough that their lifetimes are less than the time
during which the model runs, these stars also disappear). We have taken for the Jeans mass the classical expression 
$M_{J} = \frac{1}{6}\pi^{5/2}\rho^{-1/2}v_{s}^{3}G^{-3/2}$, where $\rho$ is the variable density of the gas cloud, $v_{s}$ is the sound speed, and
$G$ is the gravitational constant. If $t-t_{i}{\geq}\frac{11700}{m^{2}}$ then a star with mass $m$ formed at time $t_{i}$ is not computed at the
current time $t$.
The input free parameters of the model are: (1) the width and centre of the Gaussians (normalized to 1) for temperature, velocity component (one
for each of the three coordinates), mass, and density; (2) the ranges of values permitted for temperature, velocity component, mass, and density;
(3) the critical values adopted for temperature, velocity and density, and of course the number of time steps considered for the time
evolution.
At each time step, if the temperatures of the two colliding gas clouds are greater than the adopted critical temperature, the relative
velocity of the clouds is less than the critical velocity, and the densities of the two clouds are less than the critical density, i.e., the
collision is completely inelastic, then we assume that one cloud disappears, while the other cloud change to a new cloud with a mass equal to the sum of the masses of the initial
two clouds (the clouds merge). If all the previous conditions occur, but the relative velocity between clouds is greater than the critical velocity, then we assume that both
clouds disappear by diffusion into the diffuse ISM. If the temperatures of the two clouds are less than the critical temperature, and the two densities are greater than the
critical density, then we differentiate two cases: (1) when the relative velocity is less than the critical velocity, then star formation occurs and the two clouds disappear; and
(2) when the relative velocity is greater than the critical velocity, we assume that each of the two clouds breaks into two subclouds, for simplicity each subcloud having half of
the mass of the parent cloud. If the temperature of one cloud is less than the critical temperature and the temperature of the other is greater than the critical temperature, and
the density of the first cloud is greater than the critical density while the other density is less than the critical density, then the first cloud remains unchanged, while the
second cloud breaks into two subclouds each with a mass half the mass of the parent cloud.
All the kinematics are computed subject to momentum conservation in each kind of collision. So for the first case, i.e., a completely inelastic collision (merger), the output
velocity for each of the three dimensions is computed as $v_{i}' = \frac{M_{1}v_{i}^{1}+M_{2}v_{i}^{2}}{M_{1}+M_{2}}$, where the superscripts 1 and 2 indicate each of the two
colliding clouds, and the subscript $i$ indicates each direction in space (x, y, z). For the case in which each cloud breaks into two subclouds, we assume that for the four
resultant subclumps, those assigned 1 and 3 have the same $x, y$ velocity components computed via $v_{i}' =
\frac{2M_{2}v_{2,i}}{M_{1}+M_{2}}+\frac{(M_{1}-M_{2})v_{1,i}}{M_{1}+M_{2}}$, where $v_{j,i}$ indicates the velocity component $i$ of the initial cloud $j$, and the
$z$-direction will be such that $v_{1,z}'=-v_{3,z}'$, with $v_{1,z}'=v_{z}'$. Similarly for the other two output subclumps 2 and 4, using $v_{i}' =
\frac{2M_{1}v_{1,i}}{M_{1}+M_{2}}-\frac{(M_{1}-M_{2})v_{2,i}}{M_{1}+M_{2}}$, and $v_{2,z}'=-v_{4,z}'$, with $v_{2,z}'=v_{z}'$. For the case when one cloud remains unaltered
(called 1), while the other breaks into two pieces (called 2 and 3), the output velocity of cloud 1 changes to the $v_{i}' =
\frac{2M_{2}v_{2,i}}{M_{1}+M_{2}}+\frac{(M_{1}-M_{2})v_{1,i}}{M_{1}+M_{2}}$, while the other two components 2 and 3 have the output velocities $v_{i}' =
\frac{2M_{1}v_{1,i}}{M_{1}+M_{2}}-\frac{(M_{1}-M_{2})v_{2,i}}{M_{1}+M_{2}}$ and $v_{2,z}'=-v_{3,z}'$.
 To simulate the fact that the clouds lie on gravitationally bound orbits, we impose box shaped boundary conditions, such that the sign of the relevant velocity component
 changes when the cloud reaches the edge of the box. One can see in Fig. 3 the result compared with data.

\begin{figure}
\includegraphics[width=90mm]{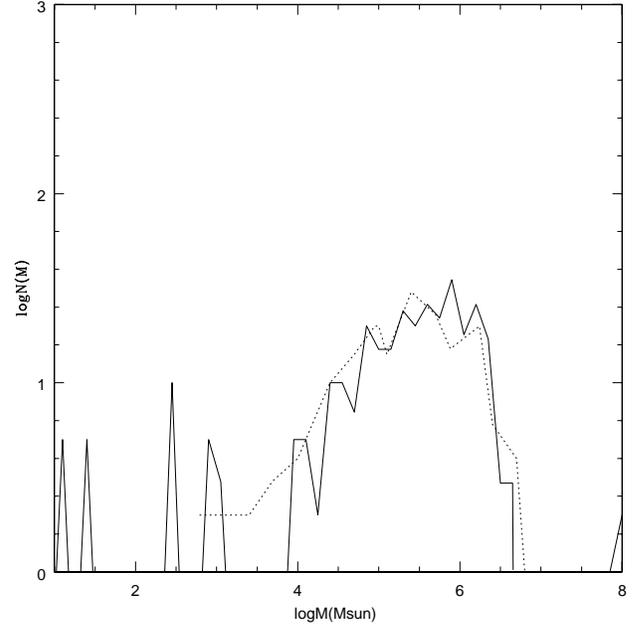}
\caption{Model of the Galactic gas cloud mass distribution function, taken from CB07, (solid line) compared with observational data from \citet{b30} dotted line.
The form of this function is not similar to that of the stellar IMFs shown in Fig. 1, but does play a role in determining the IMF as explained in Section 3 of this article.}
\end{figure}

 \subsubsection{The analytically derived molecular cloud gas mass distribution function.}
The spaces between the gas clumps inside a cloud are large, i.e., they show a low filling factor of 0.1 for clumps within giant molecular clouds (GMCs), implying a distance between
clumps of $\sim$6 pc while the size of each clump is typically of 1 pc. A similar structure is observed for the clouds in the galactic disk as a whole: characteristic
distances between clouds are of 5$\times$10$^{2}$ pc while the size of a cloud is of order 10 pc. These observational data, together with the observed relative velocities: 5 km
s$^{-1}$ between clumps and 50 km s$^{-1}$ between GMCs in the disk, permit us to make the approximation that the general distribution of gas in the ISM may be driven by
random collisions leading to a Brownian (Gaussian) distribution function of displacements, at least after a sufficiently long period of time. In fact, we can see how the
clumps of gas clouds have number distributions against size which in log-log plots appear as perfect Gaussian distributions (see CB02). Translated to linear units this implies
Planckian distributions divided by size, implying an origin based on a system formally equivalent to standing waves in a box, although in fact somewhat more complex.
As a first approximation we consider the stochastic differential Langevin equation for the time evolution of a sample of clouds under Brownian motion and
also suffering the effects of a magnetic field $B$. Assuming that $B$ is uniform, time independent, and perpendicular to the velocity $u$ of clouds, for
simplicity, one has:
\begin{equation}
\frac{du}{dt} = -{\lambda}u + A(t) - \frac{quB}{m}
\end{equation}
where $q$ and $m$ denote the charge and mass of the cloud respectively. In this equation the influence of the surrounding medium on the cloud can be split
into three terms: firstly, a systematic term $-{\lambda}u$ representing a friction experienced by the cloud, due to its movement through the other
clouds; second, a fluctuating term $A(t)$ which is the characteristic of stochastic Brownian motion; and third, the simplified influence of an assumed
uniform and constant magnetic field $B$. We have taken a magnetic force proportional to the velocity, because the magnetic force per unit volume at a
distance $R$ from the galactic centre is $\sim\frac{B^{2}}{R8\pi}$, and the magnetic flux in a galactic disk is ${\rho}u{\propto}B^{2}$. The frictional
term $-{\lambda}u$ is assumed to be governed by Stokes' law: the frictional force decelerating a spherical cloud of radius $b$ and mass $m$ is given by
$\frac{6{\pi}b{\eta}u}{m}$, where $\eta$ denotes the coefficient of viscosity of the surrounding fluid. Hence one has ${\lambda}=\frac{6{\pi}b{\eta}}{m}$.
The fluctuating term $A(t)$ has been restricted for simplicity to be independent of $u$ and with variations extremely rapid compared to the variations in
$u$.
To solve equation (3) first we write it as
\begin{equation}
\frac{du}{dt} = -{\beta}u + A(t)
\end{equation}
with
\begin{equation}
\beta = \lambda + \frac{qB}{m} = \frac{6{\pi}b{\eta}+qB}{m}
\end{equation}
Now the solution to equation (4) when $t\gg\beta^{-1}$ is:
\begin{equation}
W(x,t;x_{0},u_{0})\simeq\frac{1}{(4{\pi}Dt)^{1/2}}e^{-\frac{(x-x_{0})^{2}}{4Dt}}
\end{equation}
with $D=\frac{kT}{6{\pi}b{\eta}+qB}$ a kind of diffusion coeficient, and $W(x,t;x_{0},u_{0})$ is the probability of displacements of length $x-x_{0}$ in
any given direction at time $t$ starting from initial values $x_{0}$ and $u_{0}$. To obtain the same result from another physical point of view, we can
consider the clouds as aggregates of microclouds. Then, the probability that a microcloud arrives at a cloud surface after a path of length $x_{1}$
(assumed approximately the same as the distance between clouds) at time $t$, is $P(x_{1},t)=\frac{x_{1}}{2t(Dt\pi)^{1/2}}e^{\frac{-x_{1}^{2}}{4Dt}}$
\citet{b9}, and so, assuming that the size of microclouds is approximately constant, the distribution of sizes $x$ for the final clouds should
be $W(x,t)\propto\int_{0}^{x}P(x_{1},t)dx_{1}$, i.e., a Gaussian distribution similar to equation (6). To make the change of variable from displacements
$x$ to masses $m$, we assume that the distribution of displacements of clouds before collision with other clouds of similar size, is comparable to that
of the sizes of the clouds. This assumption, taken together with the observational relation between the sizes and masses: $x{\propto}m^{\alpha}$ with
$0.27\leq\alpha{\leq}0.41$ leads to a distribution:
\begin{equation}
\frac{dN}{dm} {\propto} m^{\alpha-1}e^{\frac{-m^{2\alpha}}{\sigma}}
\end{equation}
However, because the regions where the clouds move (mainly the spiral arms) are regions where the motion is limited spatially, we can treat the
motion formally as if were confined within a box with walls which, although not in fact fixed, may be treated as fixed to a first approximation.
And this situation can be formulated by a distribution function of displacements which is the addition of several Gaussians: the main Gaussian
(centred on zero), and the other Gaussians centred on the walls which reflect the clouds whose displacements are distributed following the main
Gaussian (see CB02). Selecting the expression for a single wall and taking $x_{0}=0$ and $u_{0}=0$ we have:
\begin{equation}
W(x,t)\simeq\frac{1}{(4{\pi}Dt)^{1/2}}\left[e^{-\frac{x^{2}}{4Dt}}+e^{-\frac{(L-x)^{2}}{4Dt}}\right]
\end{equation}
where $L/2$ is the size of the region, and the maximum of $W(x,t)$ is at $x_{0}=L/2$. We have neglected losses of clouds through the barrier
(i.e. out of the confined region). Since $x{\propto}m^{\alpha}$, one can assume $L{\propto}M_{C}^{\alpha}$, and then one has the final
distribution:
\begin{equation}
\frac{dN}{dm} {\propto} m^{\alpha-1}\left[e^{\frac{-m^{2\alpha}}{\sigma}}+e^{\frac{-(M_{C}^{\alpha}-m^{\alpha})^{2}}{\sigma}}\right]
\end{equation}

 \subsubsection{Folding the gas and dust distribution functions.}
 We propose here that there is a monotonic dependence of the mass scale of stars forming within a given gas cloud, and the scale of the dust grain
 size within the cloud. This hypothesis has $\it a$ $\it priori$ plausibility because it is based on the known requirement of dust grain surfaces to catalyze the
 formation of H$_{2}$ molecules from an initial cloud of HI. The fractal nature of the dust grains implies that the larger are the grains the greater
 is their effective catalytic surface to volume ratio, so that larger grains tend to give rise to a greater molecular fraction. There are other effects
 working in the same direction. Larger grains are formed, and maintained unfragmented, preferentially in regions with recent massive star formation,
 and as the reaction from HI to H$_{2}$ procceds more rapidly under conditions of greater hydrostatic pressure \citep{b11,b2,b3}, and supernovae arising from 
 massive stars give periodic bursts of high pressure, these two effects combine to yield more massive molecular
 clouds in regions with previous high mass stellar populations. Following from this, if high mass molecular clouds yield stars with higher ranges of
 mass, we could expect to find a relation between dust grain size and the stellar mass range, because as we have mentioned above, lower mass stars tend
 to produce dust grains of lower size, which "seed" molecular clouds of lower mass, which in their turn produce a lower range of stellar masses.
 However there is a further effect which must be taken into account, which is that the grain distribution affects the internal radiative equilibrium of
 the molecular clouds and hence their tendency to condense into cores and then stars. The external radiating surfaces of the larger dust grains are
 proportionally smaller than those of their smaller counterparts with equal volume, so that larger clouds cool less efficiently and the Jeans masses of
 their cores are higher. This effect will tend to augment the tendency of the higher mass molecular clouds to yield higher ranges of stellar
 masses. The model we use therefore has two basic inputs: the cloud mass distribution and the grain size distribution. In our first model (model
 A), we have taken a numerically derived MCGMF, from section 3.1.2 and folded it with the GSDF from section 3.1.1. In order to perform this folding we normalize the
 distributions: we take the intervals in $log r({\mu}m)$ between -3.6 and -0.4 for the GSDF (see Fig. 2) and identify these extremes with those of the MCGMF: 4 and
 6.5 in $log M_{\odot}$ (see Fig. 3). Then we divide both intervals in 100 equal parts. So, we normalize dividing each linear value of GSDF by its higher value in
 the interval, and the same for MCGMF, then transforming both distributions to other distributions normalized to unity in linear values. We then interpret the 
 normalized MCGMF and GSDF as probability densities, multiplying them together point by point, and normalizing the results to 2 in log X(log M). Finally,
 we assimilate the extremes of both intervals (that of the GSDF and that of the MCGMF) to the extremes of a new interval of stellar masses at birth taken from 0.1
 M$_{\odot}$  to 25 M$_{\odot}$. The curve we find using this semi-analytical
 approach, model A, is shown in Fig. 4 (full line). To explore an alternative approach we have used a theoretical model for the MCGMF, based on CB02, 
 in which the cloud mass distribution was computed using a simplified model in which the internal cloud dynamics was determined by the
 interaction of gravitational bounding and turbulent gas pressure. We fold this function with 
the theoretical grain size distribution of CB10 as in model A and the result is our model B (full line in Fig. 5) where we have extrapolated linearly the
results to 100 M$_\odot$.

\begin{figure}
 \includegraphics[width=90mm]{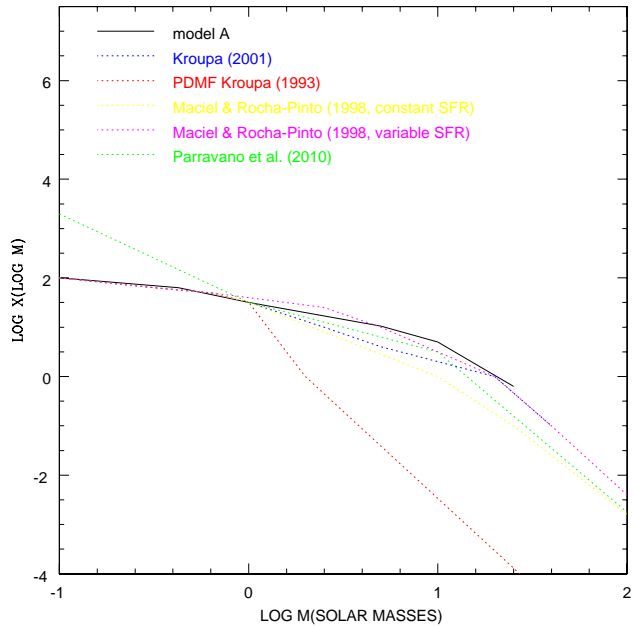}
 \caption{Prediction of our model A, using a numerical molecular cloud mass function (Section 3.1.2) folded with the dust grain size function (Section 3.1.1
 , Fig. 2), compared with three observationally based IMF's.}
\end{figure}

In the process of identification of ranges (in fact the limiting values) of cloud mass and grain size we obtain two semi-empirical relationships.
 We then make a linear fit to a statistically representative set of 100 values of $log r$ as a function of the 100 values of $log M_{C}$ and also for the 100 values of the stellar masses 
at birth $log M$ and so we obtain relations whose validity rests only on the good fits to data of the IMFs obtained from these relations (see Figs. 4 and 5). 
The first relation implies a functional dependence
between the characteristic stellar mass, $M$, formed as a result of the internal collapse of a major gas cloud and the mass, $M_{C}$, of the cloud, and is
just $M_{C}$$\sim$10$^{5}$M. The second is a relation between the characteristic stellar mass, $M$, (in units of solar mass) and the characteristic linear
size, $r$ (in $\mu$m) of the dust grains in the cloud, which is M$\sim$25r$^{0.7}$. This trend, at least in qualitative terms, can be understood from the
known requirement of grain surfaces for the catalytic conversion of HI to H$_{2}$ \citep{b12,b8,b32}. This implies that the mass of a cloud should 
increase with grain surface available. It is also known that because of the fractal geometry
of dust grains \citet{b19}, who tested grain shape models in laboratory tests) the catalytic surface varies both with radius and with
shape. Spheroidal grains have effective surfaces which increase as r$^{2}$, so their surface per unit mass falls with increasing radius, obeying an
r$^{-1}$ relation. Toroidal shapes with constant section have effective surface and volume varying as r$^{1}$, so their effective surface area per
unit mass is constant, and independent of radius. Fractal grains have effective surfaces which vary with their radii according to a power law with
index very different from 2, which would lead us to anticipate a possible dependence of cloud mass on grain size with a positive index, as we can infer from 
the result of our normalization. It is worth noting here that the
effective surface area for catalytic reactions will be, for fractal grains, much larger than the effective surface area for radiative equilibrium,
since a major proportion of the fractal surface radiates into the grain, and only the outermost surface radiates energy away from the grain.
Essentially the area to use in radiative equlibrium calculations will be approximately that of a spherical grain with the same global radius, and will thus vary
nearly as r$^{2}$ even for a fractal grain.
 
\subsection{Analytical treatment}

 Our model C is a theoretical model based on the assumption of balance between dust thermal emission and heating due to collisions
 with gas particles (\citet{b29}). This leads to the equilibrium equation:
\begin{equation}
4{\sigma}T{^{4}_{gr}}{\kappa_{P}}{\beta_{esc}}{\rho_{gr}}=n{_{gr}}2k(T_{G}-T_{gr})n_{H}{\sigma_{gr}}\left(\frac{8kT_{G}}{{\pi}m_{H}}\right)^{1/2}f
\end{equation}

where $\sigma$ is the Stefan-Boltzmann constant, $T_{gr}$ is the dust temperature, $T_{G}$ is the gas temperature, $\kappa_{P}$ is the Planck mean
opacity of dust grains, per unit mass, $\beta_{esc}$ is the photon escape probability, $\rho_{gr}$ is the dust mass
density, $n_{gr}$ is the grain number density, $\sigma_{gr}$ is the mean grain cross section, $n_{H}$ is the hydrogen number density, 
$m_{H}$ the mass of the hydrogen atom, and $k$ is the Boltzman constant. The factor $f$ takes into account the contribution to the gas-dust equilibrium
of species other than hydrogen. If we take as an approximation that the gas temperature is significantly higher than the grain temperature, we can use Eq. (2)
to derive a relatively simple relationship between the gas temperature and the grain parameters:
\begin{equation}
T_{G}^{3/2}{\propto}{\sigma_{gr}^{-1}}n{_{gr}^{-1}}
\end{equation}
As the mean grain cross section $\sigma_{gr}{\propto}r_{gr}^{2}$, and the number of grains $n_{gr}{\propto}r_{gr}^{-3}$ 
this gives us a relationship between gas temperature and grain size:
\begin{equation}
T_{G}^{3/2}{\propto}r_{gr}^{1}
\end{equation}
As the Jeans mass within a cloud is proportional to $T_{G}^{3/2}{\rho_{G}^{-1/2}}$ we find that the Jeans mass is proportional to $r_{gr}^{1}$. This
result gives a clue to the possible causal link between the characteristic dust grain size and the final characteristic stellar mass, which we show gives a
stellar mass-grain radius relationship where $M{\propto}r^{0.7}$.

\begin{figure}
 \includegraphics[width=90mm]{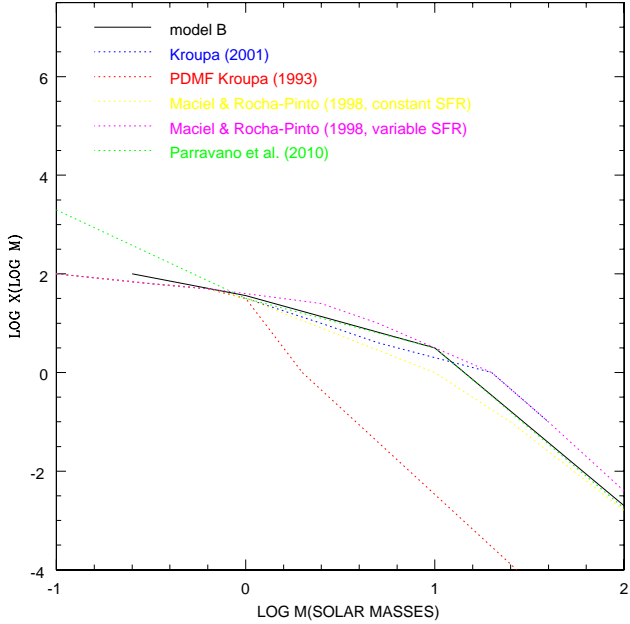}
 \caption{Prediction of our model B, which uses a theoretically derived molecular cloud mass function (Section 3.1.4) folded with the dust grain size function
 (Fig. 2), compared with the three chosen observational IMF's.}
\end{figure}

We can now go on to obtain an analytic approximation to the IMF, again assuming IMF$\propto$GSDF$\times$MCGMF, but now using an analytic fit to the data from CB10,
in the form GSDF$\propto$r$^{-3.6}$, and taking MCGMF as derived in section 3.1.3. This formulation gives us:
\begin{equation}
IMF{\propto}M^{{\alpha}-4.6}[e^{-{\frac{M^{2{\alpha}}}{\eta}}}+e^{-{\frac{(M_{C}^{\alpha}-M^{\alpha})^{2}}{\eta}}}]
\end{equation}
If we now use, following section 3.1.3, $\eta$$\propto$T$_{G}$ and starting from Eq. (12), and the relation between the Jeans mass and $r$, we find:
\begin{equation}
IMF{\propto}M^{{\alpha}-4.6}[e^{-{\frac{M^{2{\alpha}-1}}{\eta'}}}+e^{-{\frac{(M_{C}^{\alpha}-M^{\alpha})^{2}M^{-1}}{\eta'}}}]
\end{equation}
where $\alpha$, $\eta$ and $\eta'$ are constants.

\section{Comparison of the model predictions with observations}
 We can see in Fig. 1 a comparison between the classical Salpeter's IMF and the PDMF of \citet{b16} and also the IMFs of: \citet{b17}, \citet{b18} (in both cases, assuming
constant SFR and variable SFR), and \citet{b21}. In Figs. 4, 5 and 6 we can see the same data sample but compared with the predictions of our models presented in this paper (A,
B, and C respectively). Our models A and C can fit reasonably well all the data between 0.1 M$_{\odot}$ and 25 M$_{\odot}$ except that of \citet{b21} for masses lower than 1
M$_{\odot}$ where the uncertainties in the observational data are largest. But the best fit is obtained for our model B where the GSDF is numerically derived and the MCGMF is derived analitically. We are not claiming here that we are
in a position to decide finally which of the models gives the closest approach to the observations, particularly as the observations themselves are being continually
revised. Nevertheless the introduction of the grain size dependence gives fits to the predicted IMF which are significant improvements on the use of the gas cloud mass
function alone.
We can see that the model A
reproduces best the IMF derived by the later authors assuming a specific time-variable SFR produced from observational data of the
chromospheric activity of local late type dwarfs. This agreement is in some sense model-dependent, because one of the underlying
assumptions of the dust size distribution component of the model was the same time variation sequence in the SFR.

\begin{figure}
 \includegraphics[width=90mm]{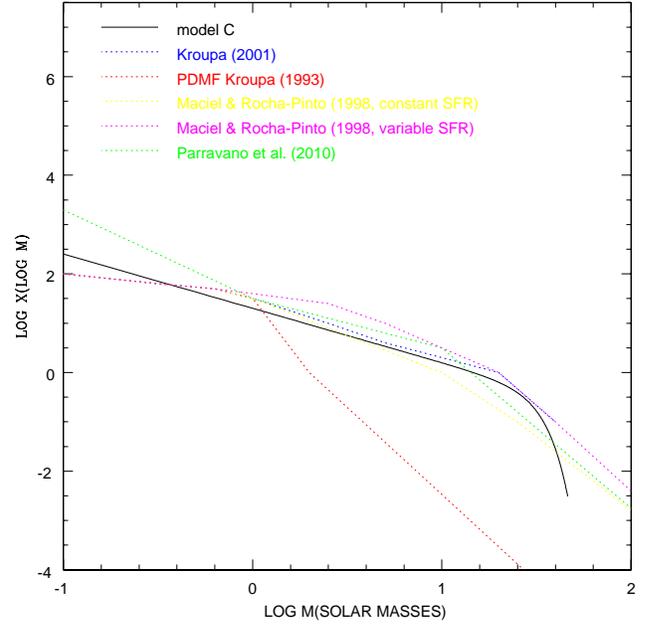}
 \caption{Prediction of our model C, in which the grain size distribution is derived from a model where the gas temperature is maintained close to equilibrium
 by radiation from the grains, which thereby influence the Jeans mass within the cloud, and hence the stellar mass. The IMF is determined using analytical expressions
 containing the cloud mass and grain size distributions. The observational IMF's are plotted for comparison.}
\end{figure}

\section{Conclusions}

This study offers prima facie evidence that there is a functional relationship between the characteristic masses of the stars at birth
and the characteristic sizes of the dust grains which populate the molecular clouds giving rise to the stars. This relationship is superposed on 
a more conventionally accepted dependence of the stellar mass range on the placental molecular cloud mass range. The evidence comes from the superior fits to the
observationally derived IMF's of the models in which the two distributions, the MCGMF and the DGSF are folded together, compared to models based on the MCGMF alone.
However we have also offered a semi-quantitative explanation based on scenarios describing the effects of the dust grains on the formation of the molecular clouds,
and on the collapse of the cloud cores to form stars. The details of these processes include two specific properties of the grains which might appear strange.
In their catalytic action leading to the formation of molecular from atomic hydrogen, the formation rates favour larger grains, because their effective surface areas are
fractal, so that their surface to volume ratio increases with grain radius. The same grains acting to radiate away heat show the opposite behaviour; their outer surfaces
from which radiation can escape are more effective for the smaller grains, because the ratio of the radiative surface to grain volume falls with grain radius. These effects
are both present in the intervention of grains in the cloud forming process and the eventual star forming process, and we have taken them quantitatively into account when
deriving the IMS's in all the models presented here. The relation of dust grain size $r$ to stellar characteristic mass $M$ can be summarized in the expression
M$\sim$25r$^{0.7}$, where $M$ is in units of solar masses, and $r$ is in microns. The dependence of the characteristic stellar mass on the mass of the placental molecular
cloud $M_{C}$ can also be parametrized, and takes the form $M_{C}$$\sim$10$^{5}$M.

\section{Acknowledgments}
 We are very grateful to the anonymous referee for the thorough comments which improved considerably the clarity in the structure of the article. This work was carried out with support from project AYA2007-67625-C02-01 of the
Spanish Ministry of Science and Innovation, and from project P3/86 of the Instituto de Astrofisica de Canarias.

\bsp

\label{lastpage}

\end{document}